\documentclass{ifacconf}
\usepackage{natbib}             
\usepackage{graphicx}          
\usepackage{amssymb}
\usepackage{mathrsfs}

\usepackage{amsmath}

\begin{document}
\begin{frontmatter}
\title{Towards Modeling HIV Long Term Behavior\thanksref{footnoteinfo}} 
\thanks[footnoteinfo]{This work was supported by Science Foundation of Ireland 07/PI/I1838, and 08/RFP/PHY/1462.}
\author[First]{Esteban A. Hernandez-Vargas}
\author[second]{Dhagash Mehta}
\author[First]{Richard H. Middleton}
\address[First]{Hamilton Institute, National University of Ireland, Maynooth, Co. Kildare, Ireland (e-mail: Richard.Middleton@nuim.ie, abelardo\_81@hotmail.com )}
\address[second]{Department of Mathematical Physics, National University of Ireland, Maynooth, Co. Kildare, Ireland (e-mail: dhagash.mehta@nuim.ie)}
\begin{abstract}
The precise mechanism that causes HIV infection to progress to AIDS is still unknown. 
This paper presents a mathematical model which is able to predict the entire trajectory of the HIV/AIDS dynamics, 
then a possible explanation for this progression is examined.
A dynamical analysis of this model reveals a set of parameters which may produce two real equilibria 
in the model. One equilibrium is stable and represents those individuals who have been living with HIV for at least 
7 to 9 years, and do not develop AIDS. The other one is unstable and represents those patients who developed 
AIDS in an average period of 10 years. However, further work is needed since the proposed model is sensitive
to parameter variations.
\end{abstract}
\end{frontmatter}
\section{Introduction} \label{Introduction}
Several mathematical models have been proposed to describe HIV dynamics since 1990, 
these present a basic relation between CD4+T cells, infected CD4+T cells and virus 
\cite{nowak00}, \cite{kirschner96}, \cite{perelson99}, \cite{xia07}. 
These models give a good presentation of the initial peak infection and the asymptomatic stage. 
However, they are not able to describe the transition to the last stage of the disease 
AIDS (acquired immunodeficiency syndrome). 

To obtain a more widely applicable model, some authors have tried to introduce other 
variables, taking into consideration other mechanisms by which HIV causes depletion of 
CD4+T cells. Numerous theories have been proposed, but none can fully explain all 
events observed to occur in practice. 
Recent studies \cite{wang00} have shown that HIV infection promotes apoptosis in
resting CD4+T cells by the homing process. This mechanism was modeled in two compartments 
by \cite{kirschner00}, in this study authors showed that therapeutic approaches 
involving inhibition of viral-induced homing and homing-induced apoptosis may 
prove beneficial for HIV patients. The role of the thymus in HIV-1 infection 
was considered by \cite{kirschner98}. The authors found that infection of the 
thymus can act as a source of both infectious virus and infected CD4+T cells. 
Significant effort has been developed in understanding the interaction of the 
immune system and HIV  \cite{fer99}, \cite{adams04}. 

One limitation of these mathematical models is that they do not reproduce 
the entire trajectory of HIV/AIDS dynamics. 
This trajectory consists of the early peak in the viral load; 
a long asymptomatic period and 
a final increase in viral load with a simultaneous collapse in healthy CD4+T cell 
count during which AIDS appears. 

A number of studies have been conducted to explore the role of macrophages in HIV 
infection as long-term reservoir \cite{orenstein01}.
A reservoir is a long-lived cell, which can 
have viral replication even after many 
years of drug treatment.  
Using this theory, \cite{conejeros07} proposed a deterministic model
which describes the complete HIV/AIDS trajectory. 
Simulations results for that model emphasize the importance of macrophages in HIV infection
and progression to AIDS, but no dynamical analysis is proposed.

In this paper, we present a simplification of \cite{conejeros07}, 
in order to have the same behavior of HIV/AIDS, which permits us 
to understand the transition to AIDS. 
The model is discussed and compared with clinical data.

\section{Model Description} \label{sec:model}
The model proposed in this section is a simplification of \cite{conejeros07}, and considers
the following populations; $T$ represents the uninfected CD4+T cells, $T_i$ represents the infected
CD4+ T cells, $M$ represents uninfected macrophages, $M_i$ represents the infected 
macrophages, and $V$ represents the HIV population. 

The mechanisms consider for this model are described
by the next reactions;
\\
\textit{A. Cell proliferation}\\
The source of new CD4+T cells and macrophages from thymus, bone marrow, and other cell sources is 
considered constant.
\begin{eqnarray}
\varnothing \stackrel{s_1}{\longrightarrow} T \label{r:s1} \\
\varnothing \stackrel{s_2}{\longrightarrow} M \label{r:s2} 
\end{eqnarray}
$s_1$ and $s_2$ are the source terms and represent the generation rate of new CD4+T 
cells and macrophages.
However, when pathogen is detected by the immune system, a signal is sent in order to become 
more aggressive, and then CD4+T cells and macrophages proliferate;
\begin{eqnarray}
T+V\stackrel{k_1}{\longrightarrow} (T+V) + T\label{r9} \\
M+V\stackrel{k_3}{\longrightarrow} (M+V) + M\label{r10} 
\end{eqnarray}
\textit{B. Infection cell}\\
HIV can infect a number of different cells; activated CD4+T cells, 
resting CD4+T cells, quiescent CD4+T cells, 
macrophages and dentritic cells. 
For simplicity, just activated CD4+T cells and macrophages are considered 
viral hosts:
\begin{eqnarray}
T+V\stackrel{k_2}{\longrightarrow} T_i \label{r1} \\
M+V\stackrel{k_4}{\longrightarrow} M_i \label{r2} 
\end{eqnarray}
\textit{C. Virus proliferation}\\
The viral proliferation is modeled as occurring in activated CD4+T cells and macrophages. 
\begin{eqnarray}
T_i\stackrel{k_5}{\longrightarrow} V + T_i\label{r4} \\
M_i\stackrel{k_{6}}{\longrightarrow} V + M_i\label{r5}
\end{eqnarray}
\textit{D. Natural death}\\
Cells and virons have a finite lifespan.
These losses are represented by the following reactions;
\begin{eqnarray}
T\stackrel{\delta_1}{\longrightarrow} \emptyset \label{r14} \\
T_i\stackrel{\delta_2}{\longrightarrow} \emptyset \label{r15} \\
M\stackrel{\delta_3}{\longrightarrow} \emptyset \label{r17} \\
M_i\stackrel{\delta_4}{\longrightarrow} \emptyset \label{r18} \\
V\stackrel{\delta_5}{\longrightarrow} \emptyset \label{r20}
\end{eqnarray}
Using reactions (\ref{r:s1})-(\ref{r20}), we obtain the following model;
\begin{eqnarray}
\dot{T}& = &s_1+k_1TV-k_2TV\nonumber - \delta_1T \nonumber \\
\dot{T_i}& = &k_2TV- \delta_2T_i\nonumber \\
\dot{M}& = &s_2+k_3MV-k_4MV-\delta_3M \label{model} \\ 
\dot{M_i}& = &k_4MV-\delta_4M_i\nonumber \\
\dot{V}& = &k_5T_i+k_{6}M_i-\delta_5V \nonumber
\end{eqnarray}
The model implementation outlined in (\ref{model}) will be conducted using MATLAB. 
Parameters and initial conditions were obtained from previous works in the area 
\cite{perelson99}, \cite{xia07}, \cite{conejeros07}. 
Using clinical data for the CD4+T cell counts \cite{greenough99}, \cite{fauci96},
some parameters were adjusted, see Table \ref{parameters}, in order to obtain 
the best match with clinical data.

\begin{table}[h]
\caption{Parameters Values}
\label{parameters}
\begin{center}
\begin{tabular}{|c|c|c|}
\hline
Parameter   & Value                                      & Value taken from:        \\      
\hline
$s_1$       & 10                                         & \cite{perelson99}        \\
\hline
$s_2$       & 0.15                                       & \cite{perelson99}        \\
\hline
$k_1$       & $2\times10^{-3}$                           & Fitted                    \\
\hline
$k_2$       & $3\times10^{-3}$                           & Fitted                     \\
\hline
$k_3$       & $7.45\times10^{-4}$                        & \cite{conejeros07}         \\
\hline
$k_4$       & $5.22\times10^{-4}$                        & \cite{conejeros07}         \\
\hline
$k_5$       & $5.37\times10^{-1}$                        & \cite{conejeros07}         \\
\hline
$k_6$       & $2.85\times10^{-1}$                        & \cite{conejeros07}         \\
\hline
$\delta_1$  & 0.01                                       & \cite{conejeros07}         \\
\hline
$\delta_2$  & 0.44                                       & Fitted                     \\
\hline
$\delta_3$  & 0.0066                                     & Fitted                     \\
\hline
$\delta_4$  & 0.0066                                     & Fitted                     \\
\hline
$\delta_5$  & 2.4                                       & \cite{xia07}         \\
\hline
\end{tabular}
\end{center}
\end{table}

\section{Model Results} \label{sec:model-results}
\begin{figure}[h]
	\centering
	\includegraphics[width=9cm,height=7cm]{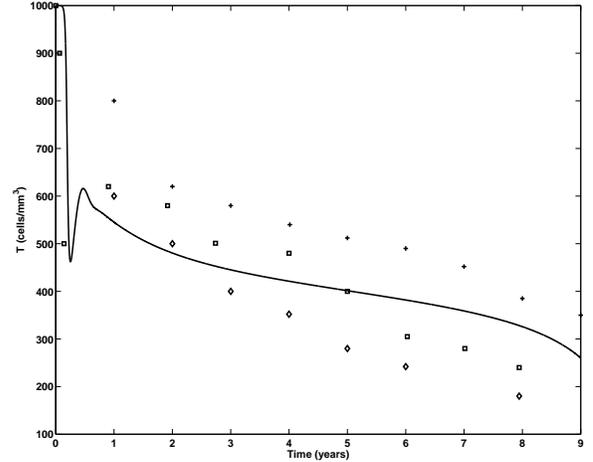}
	\caption{CD4+ T cells dynamics. Comparison with clinical data taken from \cite{greenough99} and \cite{fauci96}}
	\label{fig:Tcell}
\end{figure}

\begin{figure}[h]
	\centering
	\includegraphics[width=9cm,height=7cm]{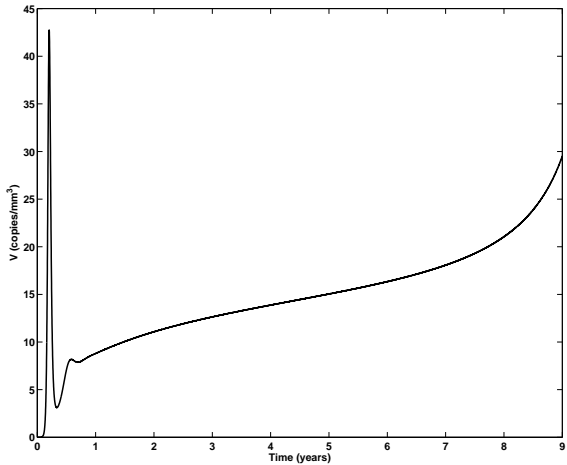}
	\caption{Viral dynamics}
	\label{fig:virus}
\end{figure}

An infected HIV  patient, in general, suffers a fast drop in healthy CD4+T cell 
count and at the same time a rapid increase in virus population. 
Then, the immune system responds to the virus by 
proliferating CD4+T cells and macrophages, this can be seen just after the dip during primary
infection in Fig.\ref{fig:Tcell}.
During the next 8 to 10 years the patient experience an asymptomatic phase. 
On one hand, CD4+T cells experience a slow depletion but are with sufficient level to
maintain most immune system functions. 
At the same time the virus population continues infecting healthy cells, 
therefore slowly advances in numbers, see Fig.\ref{fig:virus}. 
At the end of asymptomatic period, constitutional symptoms appears when CD4+T cells are 
below about 300 $mm^{-3}$. The last stage and the most dangerous for the patient is when 
CD4+T cells drop below 200 $mm^{-3}$ and the viral explosion takes place, which is considered as
AIDS. We can see in Fig.\ref{fig:Tcell} how the proposed model is able to represent all the stages 
of the infection and matches well clinical data.

\begin{figure}[h]
	\centering
	\includegraphics[width=9 cm,height=7cm]{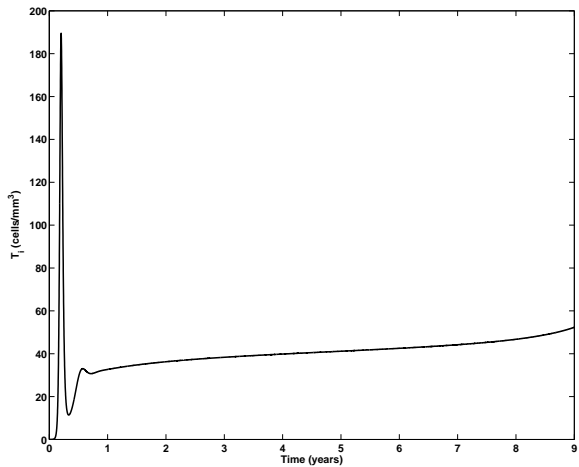}
	\caption{Infected CD4+ T cell dynamics}
	\label{fig:Ti}
\end{figure}

Infected CD4+T cell behavior is structurally similar to the viral dynamics in the first years, 
see Fig.\ref{fig:Ti}. There is an initial peak of infected CD4+T cells, followed by a small 
increment but almost constant during the asymptomatic stage. 

\begin{figure}[h]
	\centering
	\includegraphics[width=9 cm,height=7 cm]{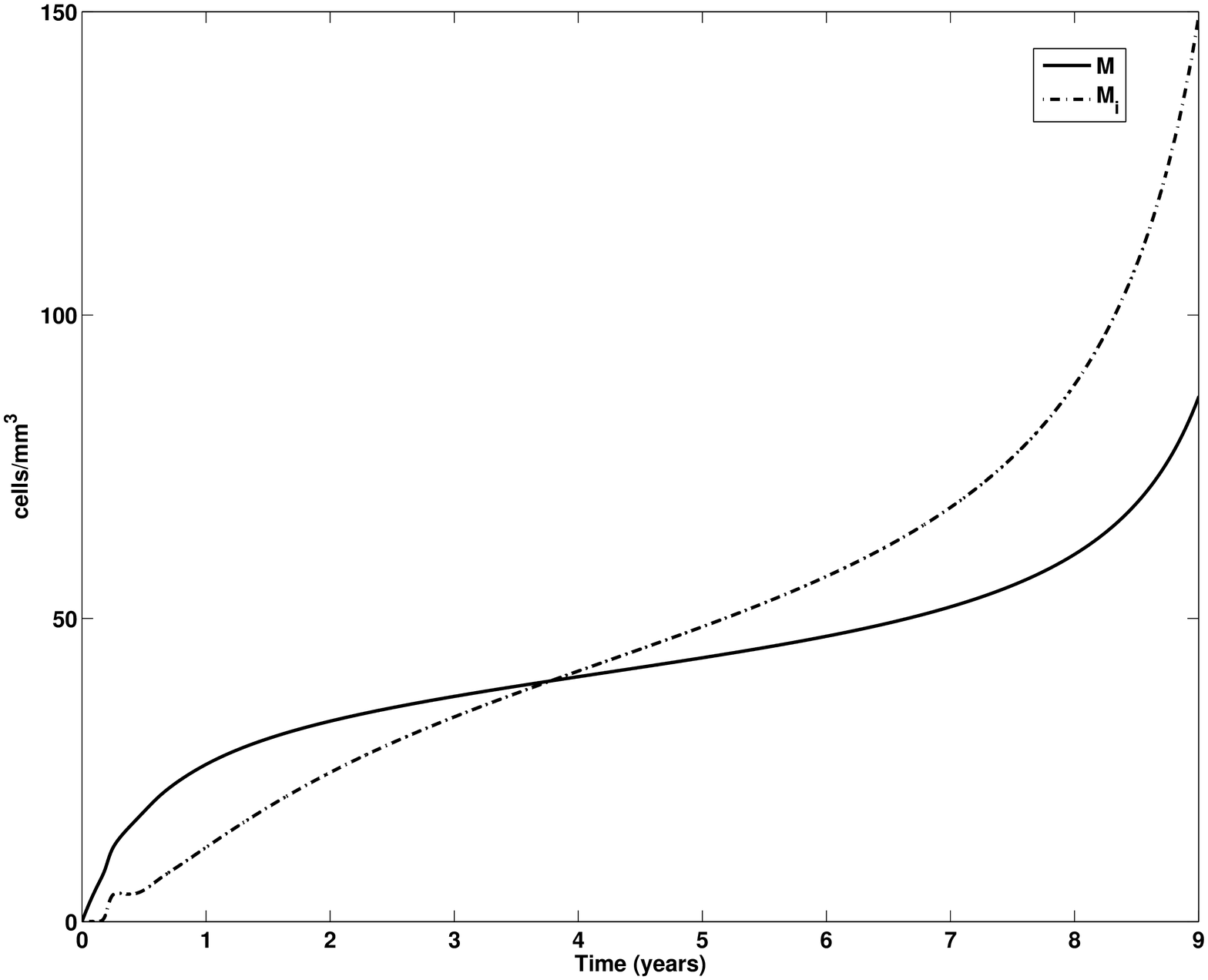}
	\caption{Macrophages and infected macrophages dynamics}
	\label{fig:M}
\end{figure}

Macrophages play a central role in this model of HIV infection. 
They are considered one of the first points of infection
and then infected macrophages are long-lived virus reservoirs as is noted in \cite{orenstein01}. 
We may notice how the model is able to represent these facts. 
Fig.\ref{fig:M} shows 
how macrophages climb over the years slowly trying to suppress the virus. This fact is 
consistent with previous works \cite{conejeros07}, who suggest that macrophages may
divide and become more aggressive.

Infected macrophages increase slowly in number during the asymptomatic period, but when 
constitutional symptoms appear infected macrophages increase very rapidly as can be seen in Fig.\ref{fig:M}. 
This is consistent with the work of \cite{igarashi01}, who argued that in the early infection the 
virus replication rate in macrophages is slower than replication rate in CD4+T cells, but over the 
period of years, the viral replication rate in macrophages is faster than early stages of infection.\\

\section{AIDS Transition} \label{sec:transition}
The model proposed by \cite{conejeros07} shows the complete HIV trajectory.
However, they do not give a detailed explanation of the HIV/AIDS transition,
since the model is difficult to analyze. 
For this reason, we propose some simplifying assumptions to allow analysis.\\
\textit{Assumption.1 Fast Viral Dynamics}\\
Looking at the differential equations \eqref{model} and parameter values in Table \ref{parameters}, 
we notice that $\delta_5>>1$. In this case, the differential equation for the virus can be 
approximated by the next algebraic equation, as noted in \cite{barao07}.
\begin{eqnarray}
V=\frac{k_5}{\delta_5}T_i+\frac{k_{6}}{\delta_5}M_i \label{Veq}
\end{eqnarray}
\textit{Assumption.2 Assume $T_i$ is bounded}\\
We note that in the asymptomatic period of infection (that is, after the initial 
transient, and before the final divergence associated with development of AIDS), 
the concentration of infected CD4+T cells is relatively constant. 
This assumption is also proposed in \cite{astolfi08}.
Therefore the following assumption for infected CD4+T cells can be considered:
\begin{eqnarray}
T_i(t) \approx \overline{T_i}, \forall t\geq t_0  \label{Tic}
\end{eqnarray}
Then under (\ref{Veq}), (\ref{Tic}) can be reduced to 
\begin{eqnarray}
V(t):=c_1M_i+V_{T_i} \label{v2}
\end{eqnarray}
where $V_{T_i}=\frac{k_5}{\delta_5}\overline{T_i}$ and $c_1=\frac{k_{6}}{\delta_5}$. 
Note that if $\bar{T}_i$ is selected as an upper bound on $T_i$, then \eqref{v2}
represents an upper bound on $v(t)$.
Therefore (\ref{v2}) describes the long asymptomatic period in the viral load dynamic.
Using last assumptions in macrophages and infected macrophages equations, 
we have the following system
\begin{eqnarray}
\dot{M} \approx s_2-c_2M+c_3MM_i \label{Meq}\\
\dot{M_i} \approx c_4M+c_5MM_i-\delta_4M_i \label{Mieq}
\end{eqnarray}
where $c_2=\delta_3-(k_3-k_4)V_{T_i}$, $c_3=(k_3-k_4)c_1$, $c_4=k_4V_{\overline{T_i}}$ and $c_5=k_4c_1$.\\ 
\textit{Assumption.3 $M$ and $M_i$ have an affine relation}\\
Note that from (\ref{Meq}) and (\ref{Mieq}), we expect that the bilinear terms are predominant 
for large $M$ and $M_i$, then we may assume $\dot{M}\approx \frac{c_3}{c_5}\dot{M}_i$, 
which are rearranged in the next linear form;
\begin{eqnarray}
M_i\approx c_6M-c_7 \label{Maprox}
\end{eqnarray}
where $c_6=\frac{c_2c_5+c_3c_4}{c_3\delta_4}$ and $c_7=\frac{c_5s_2}{c_3\delta_4}$.\\ \\

\textbf{Proposition 1.} Under Assumptions 1-3, the macrophage dynamics in an 
infected HIV patient are unstable with a finite time escape.\\
\\
\textit{Proof:}
Substituting (\ref{Maprox}) in equation (\ref{Meq}), we have a good approximation 
for the macrophage equation;
\begin{eqnarray}
\dot{M}& = &s_2+ \alpha M^2 + \beta M \label{aproxdM}
\end{eqnarray}
where
\\
$\alpha=\frac{c_6k_{6}(k3-k_4)}{\delta_5}$
\\
$\beta=\frac{(k_3-k_4)(k_{5}\overline{T_i}-c_7k_{6})}{\delta_5}-\delta_3$
\\
The solution of the differential equation (\ref{aproxdM}) for $4\alpha s_2 \geq \beta ^2$ is given by;
\begin{eqnarray}
M=\frac{\beta}{2\alpha}+\frac{\sqrt{4\alpha s_2-\beta^2}}{2\alpha} tan\left(\frac{\sqrt{4\alpha s_2-\beta^2}}{2}t+\eta \right) \label{aproxM}        
\end{eqnarray}
where $\eta$ is a constant related to initial condition of macrophages given by;
\begin{eqnarray}
\eta=tan^{-1}\left(\frac{2\alpha M_0 -\beta}{\sqrt{4\alpha s_2-\beta^2}}\right) \label{eta}                           
\end{eqnarray}
In (\ref{aproxM}) there is a tangent function, which tends to $\infty$ when the argument 
tends to $\pi/2$, that is when;
\begin{eqnarray}
t=T_\infty := \frac{\pi-2\eta}{\sqrt{4\alpha s_2-\beta^2}} \label{tiempo}                            
\end{eqnarray}
implies that there is a finite escape time.  
\begin{flushright}
$\blacksquare$
\end{flushright}

\section{Dynamical Analysis} \label{sec:analysis}
\begin{figure}[h]
	\centering
	\includegraphics[width=7cm,height=2cm]{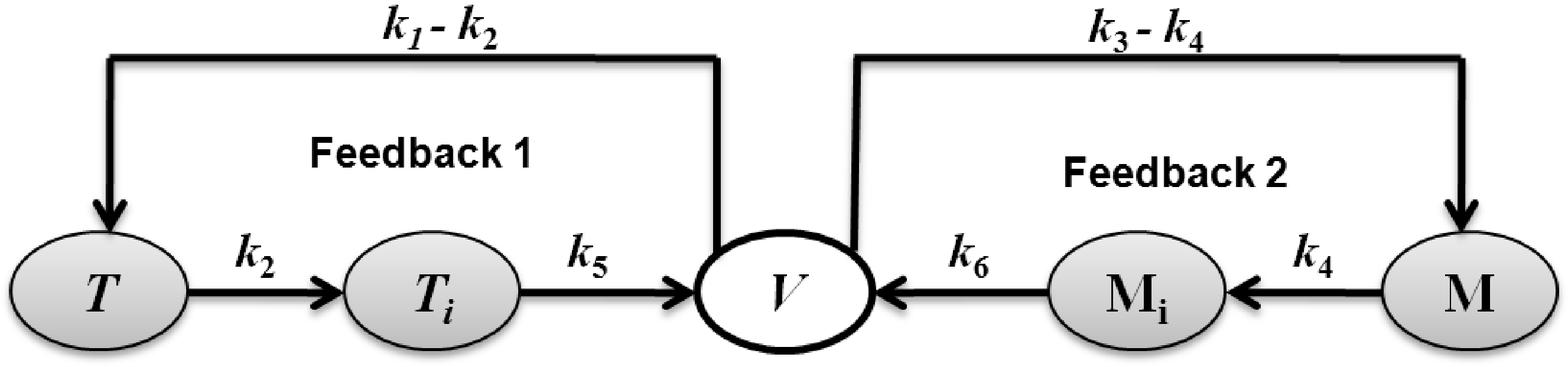}
	\caption{HIV scheme}
	\label{fig:scheme}
\end{figure}

The structure of the system (\ref{model}) can be decomposed in 
two feedbacks, see Fig.\ref{fig:scheme}; one is a fast negative feedback and the 
other is a slow positive feedback. The biological meaning of this is that feedback 1 is 
due to a fast CD4+T cell infection which is greater than the CD4+T cell proliferation. 
The feedback 2 is for a slow macrophages infection rate which is less than macrophage
proliferation. 

Using the system (\ref{model}), we are able to get the equilibrium points analytically in the form;
\begin{eqnarray}
T=\frac{s_1}{k_dV+\delta_1}   ,  T_i=\frac{k_2s_1}{\delta_2}\frac{V}{k_dV+\delta_1} \nonumber \\
M=\frac{s_2}{k_nV+\delta_3}   ,  M_i=\frac{k_4s_2}{\delta_4}\frac{V}{k_nV+\delta_3} \nonumber
\end{eqnarray}
\\
where $k_d=k_2-k_1$, $k_n=k_4-k_3$ and the value of $V$ is a solution of the polynomial.
\begin{eqnarray}
aV^3+bV^2+cV=0  \label{pol3}
\end{eqnarray}
The equation (\ref{pol3}) has three solutions, which are;
\begin{eqnarray}
V^{(A)}=0 , V^{(B)}=\frac{-b+\sqrt{b^2-4ac}}{2a}, V^{(C)}=\frac{-b-\sqrt{b^2-4ac}}{2a} \nonumber
\end{eqnarray}
where;\\
$a=\delta_2\delta_4\delta_5k_nk_d$\\
$b=\delta_2\delta_3\delta_4\delta_5k_d+\delta_1\delta_2\delta_4\delta_5k_n-\delta_4k_2k_nk_5s_1-\delta_2k_4k_dk_6s_2$\\
$c=\delta_1\delta_2\delta_3\delta_4\delta_5-\delta_3\delta_4k_2k_5s_1-k_4k_6\delta_1\delta_2s_2$\\
\\
\textit{Equilibrium A}
\begin{eqnarray}
T^{(A)}=\frac{s_1}{\delta_1},\; T_i^{(A)}=0,\; M^{(A)}=\frac{s_2}{\delta_3},\; M_i^{(A)}=0,\; V^{(A)}=0 \nonumber
\end{eqnarray}
\\
\textit{Equilibrium B,C}
\begin{eqnarray}
T^{(B,C)}=\frac{s_1}{k_1V^{(B,C)}+\delta_1}   ,\;\; T_i^{(B,C)}=\frac{k_1s_1}{\delta_2}\frac{V^{(B,C)}}{k_1V^{(B,C)}+\delta_1} \nonumber \\
M^{(B,C)}=\frac{s_2}{k_2V^{(B,C)}+\delta_3}   ,\;\;  M_i^{(B,C)}=\frac{k_2s_2}{\delta_4}\frac{V^{(B,C)}}{k_2V^{(B,C)}+\delta_3} \nonumber
\end{eqnarray}

Equilibrium A represents an uninfected status. 
Using numerical values, the uninfected equilibrium 
is unstable, which is consistent with previous works \cite{astolfi08}.
This could explain why it is difficult to revert a patient once infected,
back to the HIV-free status.
Using parameter set in Table \ref{parameters}, equilibria $B$ and $C$ 
take imaginary values since $b^2 \leq 4ac$, which are not important for the 
biological case. This means that there is no stable point for the patient 
with this set of parameters, then the model will progress to AIDS status.

\begin{figure}[h] 
\centering
\begin{tabular}{cc}
\centerline{\includegraphics[width=7cm,height=5.5cm]{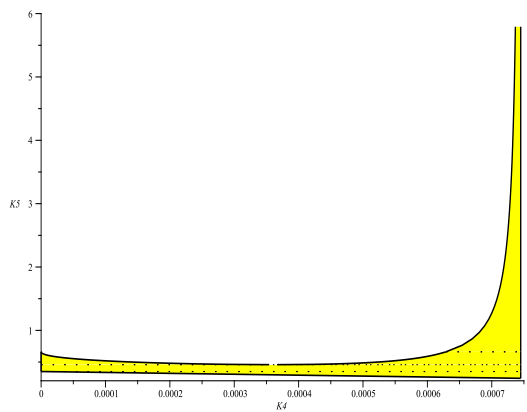}}
\end{tabular}
\caption{Parameter space in terms of the number of equilibrium points: shaded region shows two equilibrium}
\label{fig:parameters}
\end{figure}
\subsection{Number of equilibrium points and parameter spaces}
Using the recently developed symbolic real algebraic geometry methods 
we keep some parameters unfixed and get the number of equilibrium points for 
the unfixed parameter space. 
The method is called the discriminant variety method which was developed in \cite{Lazard07}, 
this is implemented in Maple as in-built packages called ``Parametric" and ``DV" \cite{Liang09}.
This method decomposes the parameter space into different cells such that the number of real roots is the 
same for any point in a given cell. We compute 2D parameter spaces below, for $k_{4}$ and $k_{5}$.
In Fig. \ref{fig:parameters} the colored region shows the part of the parameter space 
in the $k_{4}-k_{5}$ plane for which there are two equilibria in the system. 
All other parameters are set as in Table \ref{parameters}.

\textbf{Remark 1.}
Using a built-in symbolic algebra routine we confirm that
the number of equilibria with biological meaning and 
dynamic properties for the proposed model are equal to
those in \cite{conejeros07}. 
\begin{flushright}
$\blacksquare$
\end{flushright} 

\subsection{Bifurcation Analysis}
\begin{figure}[h]
	\centering
	\includegraphics[width=10cm,height=8cm]{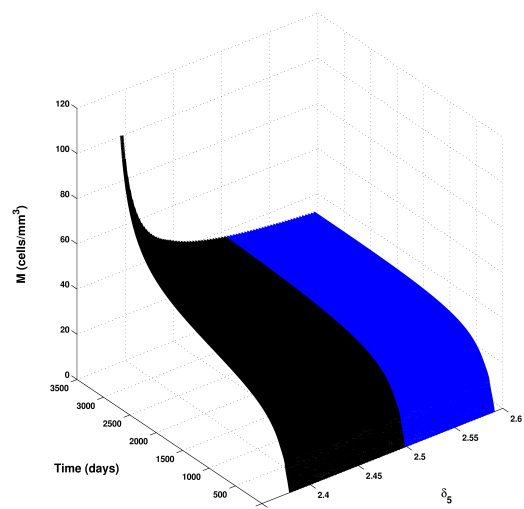}
	\caption{Bifurcation using $\delta_5$ (the black region is the unstable behavior and the blue region is stable)}
	\label{fig:d5}
\end{figure}

Using the proposed model, 
there is a set of parameters which may produce real equilibria in B and C. 
If we solve for parameter $\delta_5$, the change of stability is given by;
\begin{eqnarray}\nonumber
\delta_5>\frac{\delta_2k_5k_dk_6s_2-\delta_4k_2k_nk_5s_1}{(-\delta_1k_n+k_d\delta_3)\delta_2\delta_4}+\\  \label{cd5} \frac{2(-\delta_4k_2k_nk_5s_1\delta_2k_4k_dk_6s_2)^{\frac{1}{2}}}{(-\delta_1k_n+k_d\delta_3)\delta_2\delta_4}
\end{eqnarray}
Increasing the value of $\delta_5$ as in (\ref{cd5}), 
the escape time may be delayed.
This means that infected patients that are able to rapidly clear the virus could postpone for many years
the transition to AIDS. Furthermore, a change in the stability from unstable to stable 
is shown in Fig.\ref{fig:d5}, that is the immune system would adapt to maintain
a stable status in the patient.

\begin{table}[h]
\caption{Bifurcation}
\label{BifT}
\begin{center}
\begin{tabular}{|c|c|}
\hline
Parameter & Critical Point                              \\
\hline
$s_1$          & 50.679                                      \\
\hline
$s_2$          & 2.463                                        \\
\hline
$k_1$          & 2.95$\times10^{-3}$                          \\
\hline
$k_2$          & 3.33$\times10^{-3}$                          \\
\hline
$k_3$          & 5.337$\times10^{-3}$                          \\
\hline
$k_4$          & 5.648$\times10^{-4}$                          \\
\hline
$k_5$          & 2.721                                         \\
\hline
$k_6$          & 4.679                                         \\
\hline
$\delta_1$     & 1.178$\times10^{-2}$                          \\
\hline
$\delta_2$     & 4.732$\times10^{-1}$                           \\
\hline
$\delta_3$     & 6.99$\times10^{-3}$                            \\
\hline
$\delta_4$     & 7.4$\times10^{-3}$                            \\
\hline
$\delta_5$     & 2.497                                         \\
\hline  
\end{tabular}
\end{center}
\end{table}


\textbf{Remark 2.}
Whilst the model reproduces known long term behavior, bifurcation analysis in
Table \ref{BifT} evidences an unusually high sensitivity. 
In particular, small relative changes  in $k_2$, $k_3$, $k_4$, $\delta_1$, 
$\delta_2$, $\delta_3$, or $\delta_5$ give bifurcation to a 
qualitatively different behavior.
\begin{flushright}
$\blacksquare$
\end{flushright} 

\subsection{Numerical Results}
Choosing $\delta_5$ as stated in condition (\ref{cd5}), 
there are two real positive equilibria which are shown in Table \ref{EP}. 
One point is called LTNP (Long Term Non Progressor), because it shows the characteristic of individuals 
who have been living with HIV around 7 to 12 years and have stable CD4+T counts around 600 $cells/mm^3$ or more. 
The other point is called progressor; when CD4+T cell counts fall below 200 $cells/mm^3$, 
the patient is said to have AIDS.  

We observe in Table \ref{EV} that LTNP point is stable and the progressor point is unstable. 
The model can represent adequately both phenomena as is presented in clinical studies. 
From the bifurcation analysis, it would appear that there are certain progressor patients who
could become LNTP. However, it is unclear if this property would be preserved in more complete
models, and clinical evidence for this seems weak.
\begin{table}[h]
\caption{Equilibria}
\label{EP}
\begin{center}
\begin{tabular}{|c|c|c|c|c|c|}
\hline
Points & $V$ & $T$ & $T_i$ & $M$ & $M_i$  \\
\hline
LTNP        & 5.43& 647.99 & 24.0& 27.83 &11.95\\
\hline
Progressor   & 21.31& 319.3 & 46.40& 81.22& 136.92 \\
\hline
\end{tabular}
\end{center}
\end{table}

\begin{table}[h]
\caption{Eigenvalues for equilibria}
\label{EV}
\begin{center}
\begin{tabular}{|c|c|}
\hline
LTNP & Progressor  \\
\hline
-0.66                                                 & $-0.66\times10^{-2}$ \\
\hline
$-0.49\times10^{-1}+0.14\times10^{-1}i$               & $0.286\times10^{-2}$\\
\hline
$-0.49\times10^{-1}-0.14\times10^{-1}i$                & $-0.457\times10^{-1}$\\
\hline
-0.286                                                 & -0.238 \\
\hline
-3.35                                                  & -3.19 \\
\hline
\end{tabular}
\end{center}
\end{table}

\section{Conclusions} \label{conclusions}
The proposed model predicts the entire trajectory of the disease: initial viremia, latency, 
and the rapid increase of virus. Using this simplified model can be understood how HIV works 
in an infected patient; basically, the virus inhibits the CD4+T cell population while 
promotes the macrophages proliferation which are reservoir for virus replication. 
The long reservoir behavior of macrophages in an infected HIV patient is a possible 
explanation of why a patient progress to AIDS, and serve as recommendation for clinical study. 
However, we are concerned about the sensitivity of the model for some parameters, 
alternative mechanisms will be considered in future work.


\end{document}